\documentstyle[prd,aps,floats]{revtex}

\begin{document}

\draft
\input epsf
\twocolumn[\hsize\textwidth\columnwidth\hsize\csname
@twocolumnfalse\endcsname

\title{New type of stable Q balls 
                 in the gauge-mediated supersymmetry breaking} 

\author{S. Kasuya and M. Kawasaki}
\address{Research Center for the Early Universe, University 
  of Tokyo, Bunkyo-ku, Tokyo 113-0033, Japan}

\date{June 13, 2000}

\maketitle

\begin{abstract}
We obtain a new type of a stable Q ball in the context of
gauge-mediated supersymmetry breaking in minimal supersymmetric
standard model. It is so-calledgravity-mediation type of Q ball, but
stable against the decay into nucleons, since the energy per unit
charge is equal to gravitino mass $m_{3/2}$, which can be smaller than
nucleon mass in the gauge-mediation mechanism. We consider the
cosmological consequences in this new Q-ball scenario, and find that
this new type of the Q ball can be considered as the dark matter and
the source for the baryon number of the universe simultaneously. 
\end{abstract}

\pacs{PACS numbers: 98.80.Cq, 11.27.+d, 11.30.Fs, 95.35.+d
      \hspace{35mm} hep-ph/0006128}


\vskip2pc]

\setcounter{footnote}{1}
\renewcommand{\thefootnote}{\fnsymbol{footnote}}

The standard cosmology provides a whole description from a few minutes
after ``big bang'' to now \cite{KoTu}. One of the evidence which
supports it is the nucleosynthesis, which successfully predicts
cosmological abundances of all light elements. It requires that there
is a small asymmetry of the baryons in the universe: 
$\eta_B=n_B/n_{\gamma} \sim 10^{-10}$, where $n_B$ and $n_{\gamma}$
denote the number density of the baryon and photon, respectively. This
and other observations show that our universe is made almost entirely
of matters and devoid of antimatters. Such matter-antimatter asymmetry
is produced by baryogenesis, which takes place nonthermally through
baryon and CP violating interactions in the very early universe.

On the other hand, inflation solves many problems which cannot be
explained in the standard cosmology, such as the homogeneity,
flatness, and monopole problems. Inflation thus predicts that 
$\Omega_{{\rm tot}}= 1$, where $\Omega$ is the density parameter, the
ratio of the density to the critical density 
$\rho_c \simeq 1.9 \times 10^{-29} h^2$ g/cm$^3$, and $h$ is the
Hubble parameter normalized by 100 km/sec/Mpc. However, 
$\eta_B \sim 10^{-10}$ corresponds to $\Omega_B h^2 \simeq 0.02$, far
less than the prediction from the inflation. Even if one does not
consider the inflation, $\Omega_{{\rm matter}} \gtrsim 0.2$ is
expected from observations for dynamical properties of galaxies and
clusters of galaxies. Therefore, most of the density of the universe
has to be in the form of dark matter. 

Several mechanisms for creating baryons have been proposed, but none
of them explain directly why the universe has almost the same amount
of baryons and the dark matter. Their answer is that it is a numerical 
coincidence. However, the Q ball provides a natural scenario for
explaining both of them simultaneously \cite{KuSh,EnMc}.

In the supersymmetric standard models, Affleck-Dine (AD) mechanism
\cite{AfDi} is the most promising procedure for the baryogenesis. In
the minimal supersymmetric standard model (MSSM), there are many flat
directions consist of squarks and sleptons \cite{DiRaTh}, which can be
identified as the AD field. Its potential is almost flat but slightly
lifted up by effects of supersymmetry (SUSY) breaking. For the
mechanism of SUSY breaking, there are two famous scenarios: the
gravity- and gauge-mediated SUSY breakings. It was believed that the
AD field stays at large field value at the inflationary stage, and,
when the Hubble parameter becomes as small as the AD scalar mass after 
inflation, rolls down homogeneously its potential with rotation,
making the baryon number of the universe.

Recently, however, it was revealed that the AD field does not evolve
homogeneously, but feels spatial instabilities, which grow nonlinear
and form into Q balls~\cite{KuSh}. A Q ball is a kind of the
nontopological soliton, whose stability is guaranteed by the nonzero
charge $Q$ \cite{Coleman,Kusenko1}. In the context of the AD
baryogenesis, the charge $Q$ is the baryon number $B$. In the
gauge-mediated SUSY breaking, a Q ball is stable against the decay
into nucleons, provided that its charge is large enough so that its
energy per unit charge is less than nucleon mass: 
$E_Q/Q \simeq m_{\phi}Q^{-1/4} < 1$ GeV \cite{DvKuSh,KuSh}, where
$m_{\phi}\sim 1$ TeV, is the mass of AD field. Therefore, the Q ball
itself can be a candidate for the dark matter. On the other hand, in
the gravity-mediated SUSY breaking, the energy of a Q ball per unit
charge is essentially constant: $E_Q/Q \simeq m_{\phi} >$ 1 GeV
\cite{EnMc}. Thus, it should decay into nucleons, and the dark matter
will be lightest supersymmetric particles (neutralinos) produced in
Q-ball decays. In either case, the dark matter and the baryon number
of the universe can be explained simultaneously by the Q-ball
formation through the AD mechanism.

In all the previous studies of Q balls in the context of SUSY
breaking, the effects of gauge- and gravity-mediations are considered
separately. However, it is natural to have both effects in the
gauge-mediated SUSY breaking scenario, since the gravity-mediation
effects will dominate over the gauge ones at the large field value.
Cosmology including AD baryogenesis in such more realistic SUSY
breaking scenario was considered in Ref.~\cite{GoMoMu}. There, AD
field is regarded as a  homogeneously rotating condensate, but we
notice that it will form Q balls due to the instabilities of the
field. Particular interest is the smallness of the gravitino mass
comparing with that in the gravity-mediation scenario. It usually
ranges between 100 keV and 1 GeV. Therefore, we can imagine a new
type of a stable Q ball: the profile is the same as that in the 
gravity-mediation, but its energy per unit charge is less that 1 GeV
because of the small gravitino mass. In this Letter, we study the
cosmological consequences of Q balls (baryogenesis and the dark
matter) in the gauge-mediated SUSY breaking, taking into account the
gravity-mediation effects at large field value.

To be concrete, let us assume the following potential for the AD
field,
\begin{eqnarray}
    V(\Phi) & = & m_{\phi}^4
              \log\left(1+\frac{|\Phi|^2}{m_{\phi}^2} \right)
            \nonumber \\
            & & + m_{3/2}^2|\Phi|^2 \left[ 1 + K\log \left(
                 \frac{|\Phi|^2}{M_*^2}\right)\right],
\end{eqnarray}
where $m_{3/2}$ is the gravitino mass, $K(<0)$ term a one-loop
correction,
\footnote{Usually, gaugino contributions to $K$ is dominated and $K$
becomes negative. However, if $K>0$ by some large Yukawa couplings,
the AD field can be stabilized and Q balls can be created only for
$\phi \lesssim \phi_{{\rm eq}}$, where $\phi_{{\rm eq}}$ is defined in 
Eq.(\ref{eq}).}
$M_*$ the renormalization scale, and we assume that the second term
should be neglected for small field value. This is nothing but the sum
of the potentials for the gauge- and gravity-mediation mechanisms
studied previously \cite{KuSh,EnMc,KK1,KK2}. However, as
we mentioned earlier, the gravitino mass is considerably smaller. The
second term will dominate the potential when 
\vspace{-0.5mm}
\begin{equation}
    \label{eq}
    \phi \gtrsim \phi_{{\rm eq}} 
                 \equiv \sqrt{2}\frac{m_{\phi}^2}{m_{3/2}},
\end{equation}
where $\Phi=\phi\exp(i\omega t)/\sqrt{2}$. In this case, a new type of
a stable Q ball is produced. Its property is very similar to that in
the gravity-mediation, such as \cite{EnMc,KK3}
\begin{eqnarray}
    R_Q \simeq |K|^{-1/2}m_{3/2}^{-1}, 
    & \qquad & \omega \simeq m_{3/2}, \nonumber \\
    \phi \simeq |K|^{3/4}m_{3/2}Q^{1/2}, 
    & \qquad & E_Q \simeq m_{3/2} Q,
    \label{gravity}
\end{eqnarray}
but, as can be seen from the last equation,  it is stable against the
decay into nucleons. In the opposite case, the Q-ball properties are
the same as in the gauge-mediation only \cite{KuSh}:
\begin{eqnarray}
    R_Q \simeq m_{\phi}^{-1}Q^{1/4},
    & \qquad & \omega \simeq m_{\phi}Q^{-1/4}, \nonumber \\
    \phi \simeq m_{\phi}Q^{1/4}, 
    & \qquad & E_Q \simeq m_{\phi} Q^{3/4}.
    \label{gauge}
\end{eqnarray}
The energy per unit charge can be treated from unified viewpoint. The
largest charge of the Q ball formed depends linearly on the initial
charge density of the AD field as~\cite{KK1,KK2} 
\begin{equation}
    Q \simeq \beta \frac{q(0)}{m_{3/2}^3} 
    \simeq \beta \left(\frac{\phi(0)}{m_{3/2}}\right)^2,
\end{equation}
where $\beta \lesssim 1$ \cite{KK3}, and we use 
$q=\omega\phi^2\simeq m_{3/2}\phi^2$. It can be understood by
estimating Q-ball charge in the standard way. The charge is given by 
\begin{equation}
    Q = \int d^3 x \omega \phi^2 
    = \left(\frac{\pi}{2}\right)^{3/2}\omega \phi_0^2 R^3
    = \beta' \left( \frac{\phi_0}{m_{3/2}}\right)^2,
\end{equation}
where we assume the Gaussian profile ansatz \cite{EnMc}, 
$\phi = \phi_0 \exp(-r^2/R^2)$, which is a very good approximation,
and use relations (\ref{gravity}), and 
$\beta' \simeq 2\times 10^3 (|K|/0.01)^{-3/2}$.
The discrepancy of coefficients $\beta$ and $\beta'$ comes from the
fact that $\phi(0)\neq\phi_0$ and there are more than one Q balls with 
charges of the same order of magnitude as the largest.

\begin{figure}[t!]
\centering
\hspace*{-7mm}
\leavevmode\epsfysize=5cm \epsfbox{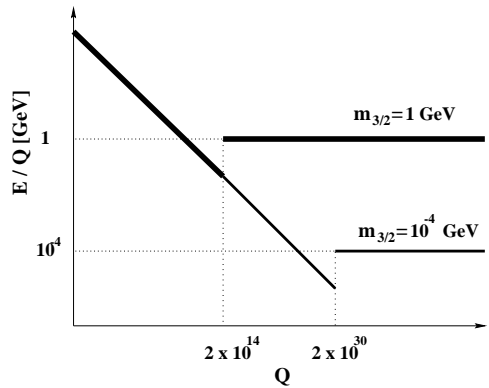}\\[2mm]
\caption[fig-1]{\label{e-q}
Dependence of the energy per unit charge on the charge of the Q ball. 
Gauge-mediation type of Q ball are formed with less charge, while
gravity-mediation type has larger with $E/Q$ fixed.}
\end{figure}

Inserting it into Eq.(\ref{eq}), we obtain
\begin{equation}
    \label{mgravitino}
    m_{3/2} \gtrsim (2\beta)^{1/4} m_{\phi}Q^{-1/4}.
\end{equation}
The right hand side of this equation is identical to the expression
for the energy per unit charge of the gauge-mediation besides the
factor of order unity. The energy per unit charge of the Q ball is
written as
\begin{equation}
\frac{E_Q}{Q} = 
\left\{
\begin{array}{ll}
m_{\phi}Q^{-1/4} & \qquad \phi \lesssim \phi_{{\rm eq}} \\[2mm]
m_{3/2} &  \qquad \phi \gtrsim \phi_{{\rm eq}}
\end{array}
\right. \,,
\end{equation}
which is shown in Fig.~\ref{e-q}. The gap on the boundary should
disappear and both sides of the curves will be smoothly connected
because Q balls formed in this region are not the exact type of either
(\ref{gravity}) or (\ref{gauge}), but will show properties between
them.

Since Q balls are stable even for $\phi \gtrsim \phi_{{\rm eq}}$,
where the gravity-mediation effect is crucial, the baryon number in
the universe should be explained by the baryons evaporated from Q
balls, as is the same as for the gauge-mediation type \cite{LaSh}. 
The evaporation rate of the Q ball is \cite{LaSh}
\begin{equation}
    \label{evap}
    \Gamma_{{\rm evap}}\equiv\frac{dQ}{dt}
                         =-\alpha\mu T^2 4\pi R_Q^2,
\end{equation}
where $\mu$ is a chemical potential of the Q ball, which is estimated
as $\mu \simeq \omega$ because $\omega$ is energy of $\phi$-field
inside the Q ball. Although the mass of the (free) AD particle
$m_{\phi}$ is affected by thermal corrections, which should be changed
as $m_{\phi} \rightarrow m_{\phi}(T) \sim T$, at $T\gg m_{\phi}$, the
gravitino mass is not affected, since particles coupled to the AD
field are decoupled from thermal bath when the AD field has a large
vacuum expectation value. At $T\gtrsim m_{\phi}$, large numbers of
$\phi$-particles are in thermal bath outside Q balls, so $\alpha \sim
1$. On the other hand, since only light quarks are in thermal bath at
$T\lesssim m_{\phi}$, the corresponding cross section is highly
suppressed by the heavy gluino exchanges, and 
$\alpha \simeq (T/m_{\phi})^2$. 

However, if the rate of the charge diffusion from the ``atmosphere''
of the Q ball is smaller than the evaporation rate, chemical
equilibrium will established there, which results in the
suppression of the evaporation~\cite{BaJe}. The diffusion rate is
\cite{BaJe}
\begin{equation}
    \label{diff}
    \Gamma_{{\rm diff}}\equiv\frac{dQ}{dt}
                        =-4\pi\zeta R_Q D \mu_Q T^2,
\end{equation}
where $D=a/T$ with $a\simeq 4-6$, and $\zeta \sim 1$. 

The time scale of charge transportation is determined by the diffusion 
when $\Gamma_{{\rm diff}}\lesssim\Gamma_{{\rm evap}}$. It holds for
$T\gtrsim T_* \equiv a^{1/3}|K|^{1/6}(m_{3/2}m_{\phi}^2)^{1/3}$. In
this case, using Eqs.(\ref{gravity}) and (\ref{diff}), and assuming
the radiation-dominated universe, $t=AM/T^2$, where $A\approx 0.2$ 
and $M\simeq 2.4\times 10^{18}$ GeV, we obtain 
\begin{equation}
    \label{high-T}
    \frac{dQ}{dT} = \frac{8\pi aAM}{|K|^{1/2}T^2}.
\end{equation}

On the other hand, when $T\lesssim T_*$, the diffusion effect is
negligible, and Eq.(\ref{high-T}) should be replaced by
\begin{equation}
    \frac{dQ}{dT} = \frac{8\pi AMT}{|K|m_{3/2}m_{\phi}^2}.
\end{equation}
Therefore, total amount of the charge evaporated from the Q ball is
\begin{equation}
    \Delta Q \simeq 12\pi AM
        \left( \frac{a}{|K|} \right)^{2/3}
        ( m_{3/2}m_{\phi}^2)^{-1/3}.
\end{equation}

Provided that the initial charge of the Q ball is larger than the
evaporated charge, we regard that the Q ball survives from
evaporation, and contributes to the dark matter of the universe:
\begin{equation}
    \label{survive}
    Q_{{\rm init}} \gtrsim \Delta Q 
    \simeq
        9.8\times 10^{18} 
        \left( \frac{m_{3/2}}{{\rm GeV}} \right)^{-1/3}
        \left( \frac{m_{\phi}}{{\rm TeV}} \right)^{-2/3},
\end{equation}
where we set $a=4$ and $|K|=0.01$.

Now we can relate the baryon number and the amount of the dark matter
in the universe. As mentioned above, the baryon number of the universe
should be explained by the amount of the charge evaporated from Q
balls, $\Delta Q$, and the survived Q balls become the dark matter. If 
we assume that Q balls do not exceed the critical density of the
universe, i.e., $\Omega_Q \lesssim 1$, and the baryon-to-photon ratio
as $\eta_B \sim 10^{-10}$,  
\begin{equation}
    \label{dm}
    Q \lesssim 3.2h^2\times 10^{21}
        \left( \frac{m_{3/2}}{{\rm GeV}} \right)^{-4/3}
        \left( \frac{m_{\phi}}{{\rm TeV}} \right)^{-2/3}.
\end{equation}

\begin{figure}[t!]
\centering
\hspace*{-7mm}
\leavevmode\epsfysize=5.5cm \epsfbox{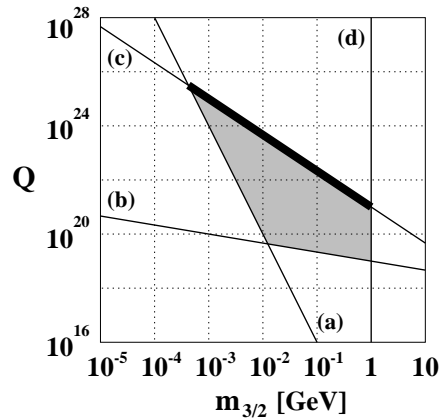}\\[2mm]
\caption[fig-2]{\label{Q-m} 
Summary of constraints on ($Q,m_{3/2}$) plain for $|K|=0.01$. (a),
(b), and (c) represent for the constraints (\ref{Q-phi-eq}),
(\ref{survive}) and (\ref{dm}) with  $m_{\phi}=1$  TeV, respectively.  
(d) shows the condition for the gravity-mediation type of the Q ball
to be stable.}
\end{figure}

Rewriting Eq.(\ref{mgravitino}), we have
\begin{equation}
    \label{Q-phi-eq}
    Q \gtrsim 2\beta \times 10^{12} 
      \left( \frac{m_{\phi}}{{\rm TeV}} \right)^4
      \left( \frac{m_{3/2}}{{\rm GeV}} \right)^{-4}.
\end{equation}
Combining this constraint with Eqs.(\ref{survive}) and (\ref{dm}),
together with $m_{3/2} \lesssim 1$ GeV, which implies that the
gravity-mediation type of the Q ball is stable against the decay into
nucleons, we obtain the allowed region for the new type of the stable
Q ball explaining the baryon number of the universe. Figure \ref{Q-m}
shows the allowed region on $(Q,m_{3/2})$ plane for $m_{\phi}=1$
TeV. The shaded regions represent that the new type of stable Q balls
are created, and the baryon number of the universe can be explained by
the mechanism mentioned above. Furthermore, the new type of stable Q
balls contribute crucially to the dark matter of the universe at
present, if the Q balls have the charge given by the thick line in the
figure. Notice that Q balls with very large charge are not allowed
because they will overclose the density of the universe.~\footnote{
If the baryons are produced by other mechanism, larger Q
balls can be allowed. In this case, however, a nice relation between
the densities of the baryon and the dark matter does not hold.}
Therefore, the initial conditions for the AD field is restricted
severely~\cite{KK3}. 

One may wonder if these new type of stable Q balls can be detected.
When a Q ball collides with nucleons, they enter the surface layer of
the Q ball, and dissociate into quarks, which are converted into
squarks. In this process, Q balls release $\sim 1$ GeV energy per
collision by emitting soft pions. This process is the basis for the
Q-ball detections \cite{KuKuShTi,ArYoNaOg}, which is called
(Kusenko-Kuzmin-Shaposhnikov-Tinyakov) KKST process in the
literature. It occurs for electrically neutral Q balls (ENQB). For
electrically positively charged Q balls (EPCQB), the KKST process is
strongly suppressed by Coulomb repulsion, and only electromagnetic
processes will take place. For electrically negatively charged Q balls
(ENCQB), the both KKST and electromagnetic processes occur, but the
former is dominant, which is essentially the same as for ENQBs. 

\begin{figure}[t!]
\centering
\hspace*{-7mm}
\leavevmode\epsfysize=7.4cm \epsfbox{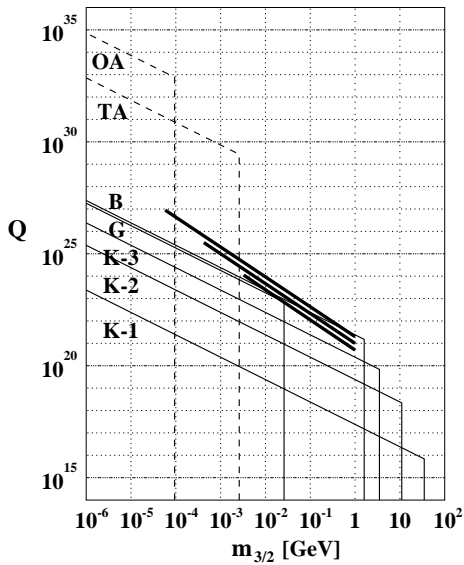}\\[2mm]
\caption[fig-2]{\label{SENS} 
Restrictions of ENQBs by several experiments on ($Q,m_{3/2}$) plain
for $|K|=0.01$. We show the regions currently excluded by BAKSAN (B),
Gyrlyand (G), and Kamiokande (K-1, K-2, K-3), and to be searched by
the Telescope Array Project (TA) and OWL-AIRWATCH (OA) in the future. 
The thick lines represent for the gravity-mediation type of the Q ball
to be both the dark matter and the source for the baryons of the
universe for $m_{\phi}=300$ GeV, 1 TeV, and 3 TeV from the top to the
bottom.}
\end{figure}

In either case, the detection is more difficult than for the
gauge-mediation type of Q balls, since the geometrical cross section
is smaller for large $Q$, and the Q-ball mass is larger for the same
$Q$, which results in small flux. With the discussions similar to
Ref.~\cite{ArYoNaOg}, we can restrict the parameter space 
$(Q,m_{3/2})$ by the several experiments. Fig.~\ref{SENS} shows the
results for ENQBs. Lower left regions are excluded by the various
experiments. The present available data prohibit the stable
gravity-mediation type Q balls with large gravitino mass to be both
the dark matter and the source of the baryons, and future experiments
such as the Telescope Array Project or the OWL-AIRWATCH detector may
detect this type of Q balls with an interesting gravitino  mass 
$\sim 100$ keV.

For EPCQBs with $Z=1$, similar constraints as for ENQBs are put by the 
MACRO experiment \cite{KK3}.

In summary, we have obtained a new type of a stable Q ball in the
context of gauge-mediated SUSY breaking in MSSM. Many properties are
the same as the gravity-mediation type of Q ball, but it is stable
against the decay into nucleons, since the energy per unit charge is
equal to the gravitino mass $m_{3/2}$, which can be smaller than
nucleon mass of 1 GeV in the gauge-mediation mechanism. We have
considered the cosmological consequences in this new Q-ball
scenario. Because of its stability, it can be a nice candidate for the
dark matter of the universe. In the present case, the baryons are
produced only by evaporation from Q balls, since (almost) all the
baryons are trapped in Q balls during their formation. We have found
that the Q ball with $Q \sim 10^{25}-10^{22}$ can account for both
the dark matter and the baryon number of the present universe for 
$m_{3/2}\simeq 10^{-4}-10^{-1}$ GeV and $m_{\phi}=1$ TeV, and such Q
balls may be detected by the future experiments.  

The authors are grateful to R. Banerjee for useful comments and
T. Yoshida for helpful discussions. M.K. is supported in part by the
Grant-in-Aid, Priority Area ``Supersymmetry and Unified Theory of
Elementary Particles''($\#707$).

\end{document}